\pdfoutput=1
\documentclass[
amsmath,amssymb,
aps,
pre,
]{revtex4-2}
\usepackage{xcolor,ulem}
\usepackage{graphicx}
\usepackage{dcolumn}
\usepackage{bm}
\usepackage{xr}
\begin{document}

\title{Promoting collective intelligence: The advantage of temporal star-structures}

\author{Zhenglong Tian$^{1,\dagger}$}
\author{Yao Meng$^{1,\dagger}$}%
\author{Wenxuan Fang$^1$}
\author{Aming Li$^{1,2,*}$}
\affiliation{%
\rm
$1$ Center for Systems and Control, College of Engineering, Peking University, Beijing 100871, China\\
$2$ Center for Multi-Agent Research, Institute for Artificial Intelligence, Peking University, Beijing 100871, China\\
${\dagger}$ These authors contributed equally to this work\\
$*$ Corresponding author. E-mail: amingli@pku.edu.cn
}%
\date{\today}

\begin{abstract}
    System structures play an essential role in the emergence of collective intelligence in many natural and engineering systems. In empirical systems, interactions among multiple agents may change over time, forming a temporal network structure, where nodes represent the system's components and links capture who interacts with whom. Recent studies report that temporal networks are more conducive to the emergence of collective cooperation compared to their aggregated static structures. However, the question of which kind of structural characteristics of temporal networks promote collective cooperation still remains elusive. Here we systematically investigate the evolution of cooperation on temporal networks with diverse structural characteristics, such as random, star, and cluster structures. We uncover that temporal networks with single-star structures which lack network clusters are more conducive to collective cooperation than other structures. This counterintuitive result cautions against the common belief that network clusters normally facilitate collective cooperation, revealing the unique advantages of temporal networks over static networks. We further propose an index to quantify the capacity of structural characteristics of temporal networks in promoting collective cooperation. Our findings pave the way for designing the optimal structure of temporal networks to favour collective cooperation.
\end{abstract}

\maketitle
    \section{INTRODUCTION}
    Collective intelligence abounds in the real world \cite{10.1093/nsr/nwad040}, emerging from the complex dynamics of natural interaction systems including social networks \cite{axelrod1981evolution,ohtsuki2006simple,nowak1992evolutionary,hauert2004spatial,allen2017evolutionary,traulsen2005coevolutionary,roca2006time,wang2023imitation,santos2006evolutionary,meng2023impact}, biological swarms, and ecosystems \cite{may1972will}, as well as autonomous coordination of machine interaction systems like distributed computing, traffic networks, and environmental monitoring systems.
    Under the canonical Prisoner's Dilemma, individual incentives are misaligned with the collective incentives of an entire group. This further leads to the extinction of collective cooperation {of the group}. Indeed, in well-mixed populations, individuals tend to adopt defection, which is the Nash equilibrium for the Prisoner's Dilemma. Interestingly, the introduction of network structures can promote the emergence of collective {intelligence. Namely, {individuals give up their individual optimal choices, leading the system to exhibit an emergent group-optimal outcome} --- collective cooperation}.
    It is well known that on complex networks where nodes represent individuals and links capture the interactions among whom, cooperators can {form cooperative clusters, where mutual cooperation within the cluster gains high payoffs, which exceed the temptation to defect, effectively resisting the invasion of external defectors.}

    However, these studies generally rely on a crucial assumption that the network structures are static. In empirical scenarios, many network structures are time-varying or stochastic, resulting in prominently different behavioural dynamics \cite{pinheiro2016linking, helbing2009outbreak,sheng2023constructing}. 
    For instance, in computer networks, interactions between devices may occur in different time periods. Similarly, in control systems, signal-carrying paths usually change over time \cite{li2017fundamental}. 
    Although recent studies suggest that such temporal networks---wherein the structure varies over time---actually enhance the evolution of cooperation compared to static structures \cite{li2020evolution}, it remains an open problem about which specific types of {structural characteristics of temporal networks} most enhance collective cooperation. In light of this, we construct five different types of temporal networks with different structural characteristics. Surprisingly, we find that a star-structure network, which has an extremely low clustering coefficient, exhibits a prominent advantage in fostering cooperation. We uncover the mechanisms behind this phenomenon and propose a simple index to evaluate the capacity of network structure in promoting collective cooperation. Our results pave the way to facilitate the emergence of cooperation by designing a structural temporal network, on a designed sequence of activated nodes and edges.

    \section{MODEL}
        We conduct an investigation within the framework of classic evolutionary game theory  \cite{nowak1992evolutionary, santos2006graph}, where each player chooses cooperation (C) or defection (D). When both players choose to cooperate (defect), they receive a payoff of $R$ ($P$). When players choose different strategies, the defector gains a higher payoff $T$ and the cooperator a lower payoff $S$. The outcomes can be expressed as a payoff matrix: 
        
        $$\begin{array}{c|cc}&\mathrm{C}&\mathrm{D}\\\hline\mathrm{C}&R&S\\\mathrm{D}&T&P\end{array} .$$Following the canonical practice \cite{nowak1992evolutionary,li2020evolution}, we investigate the (weak) Prisoner's Dilemma as $R=1$, $P=S=0$, and $T=b>1$. Here, $b$ is the only parameter that captures the temptation of defection.

        \begin{figure*}[htbp]
            \centering
            \includegraphics[width=\linewidth]{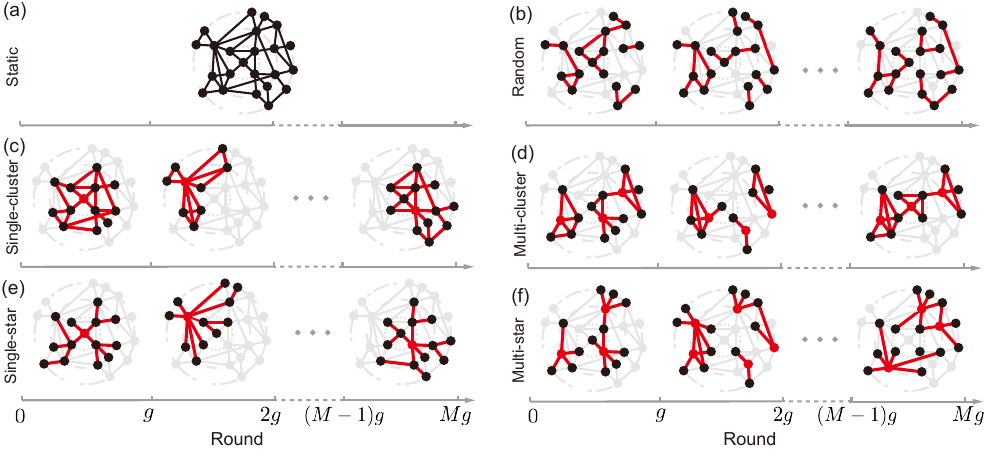}
            \caption{Illustration of static scale-free networks and the corresponding five structural temporal networks with different structural characteristics. (a) The static scale-free network---wherein the edges do not vary over time---is the basis of the generated temporal networks.
            The temporal interactions between individuals in a population are then captured by a sequence of $M$ snapshots, where black (grey) nodes indicate active (inactive) individuals and red (grey) edge captures the active (inactive) interactions at each snapshot.
            We construct structural temporal networks by activating nodes and edges from the underlying static scale-free networks following specific rules in each snapshot.
            ({b}) First, we generate the random temporal network by randomly choosing edges from the underlying static network in ({a}) with proportion $p$. 
            ({c}) We generate the single-cluster structure on each snapshot independently by selecting edges according to the Breadth-first search (BFS) method from a random starting node (red). 
            ({d}) The only difference between a single-cluster and a multi-cluster structure arises from the search levels (marked in grey dashed circles) from the initial starting node, with the multi-cluster structure only allowing to connect the 1st level of neighbours. Note that to satisfy the requirement for a sufficient number of edges, this structure will have multiple initial nodes (red).
            ({e}) We generate the single-star structure by performing the spanning tree algorithm from a random starting node (red), which enables to avoid the formation of triangle structures during the construction.
            ({f}) Similarly, the multi-star structure can only be generated with 1st level of neighbours.
            For the evolutionary game process on the temporal networks, we run $g$ rounds of games on each snapshot before switching to the next, and a total generation $G$ is evolved during the process. If $Mg<G$, {we restart the sequence of snapshots from the beginning.}
            }\label{Fig1}
        \end{figure*}

        The evolutionary game process occurs on a sequence of subnetworks (snapshots) formed by selecting edges on the underlying static network (Fig.~\ref{Fig1}). Initially, each player on the first snapshot chooses cooperation or defection with equal probability. In each round of the game, each player $i$ plays the game with all of its neighbours and obtains the accumulated payoff $P_i$. Afterwards, player $i$ randomly chooses one of its neighbours $j$ and adopts the strategy of $j$ with probability $p_{i\rightarrow j}=1/(1+\exp{[-s(P_j-P_i)]})$, where $s>0$ captures the selection intensity \cite{perc2006coherence,perc2010coevolutionary}. The population evolves for $g$ rounds in each snapshot before switching to the next one, until a given time or one of the strategies goes extinct. In this way, $g$ denotes the total number of interactions between players in a snapshot, namely the relative time-scale between dynamics of game and network structure.

        We explore the evolution of cooperation on temporal networks by generating the network with five different typical characteristics reported in real-world empirical systems. Figure \ref{Fig1}a shows a static scale-free network constructed by Barab{\'a}si-Albert model  \cite{barabasi1999emergence}, where the structure does not change over time (see Supplementary Fig.~S1 for a snapshot). We construct five types of temporal networks, each characterized by a different method of selecting edges from the underlying network, with a fraction of selected edges $p$. The random temporal networks, which we take as a null model, are constructed by randomly selecting edges from the static network to form each snapshot (Fig.~\ref{Fig1}b, also see Supplementary Fig.~S2 for snapshots and Supplementary Algorithm S1 for the pseudocode). 
        
        To explore how structural characteristics affect the evolution of cooperation, 
        we construct temporal networks based on two typical structures---clusters and stars.
        The first type is the single-cluster temporal network, which is constructed by first including edges between an initial random node and its neighbours, then the edges between the neighbouring nodes, and further repeating this procedure on the neighbouring nodes (Fig.~\ref{Fig1}c, see Supplementary Fig.~S3 and Supplementary Algorithm S2 for details). 
        Similarly, we construct the multi-cluster temporal network, which starts from multiple random nodes and only expands to {their} first order neighbours (Fig.~\ref{Fig1}d, see Supplementary Fig.~S4 and Supplementary Algorithm S3 for details).
        {Furthermore}, we {also} construct structural temporal networks based on star structures. 
        The single-star temporal network is composed {of} snapshots wherein the structure grows with an initial randomly selected node, involving all its neighbours, and then expands by adding edges to neighbours not in the structure (Fig.~\ref{Fig1}e, see Supplementary Fig.~S5 and Supplementary Algorithm S4 for details).
        Similarly, the multi-star structures only expand to first order neighbours compared to the single-star structure (Fig.~\ref{Fig1}f, see Supplementary Fig.~S6 and Supplementary Algorithm S5 for details).
        These structures are designed to represent a broad range of real-world networked systems. For instance, the single-cluster and multi-cluster structures can reflect social networks where communities or groups form around single or multiple influential individuals or topics. On the other hand, the single-star and multi-star structures can represent hierarchical systems such as company organizational structures or web page links, where a central node (like a home page) connects to many other nodes.

    \section{RESULTS}
        \subsection{Characteristics of temporal networks significantly alter the evolution of cooperation}

\begin{figure*}[htbp]
                \centering
                \includegraphics[width=\linewidth]{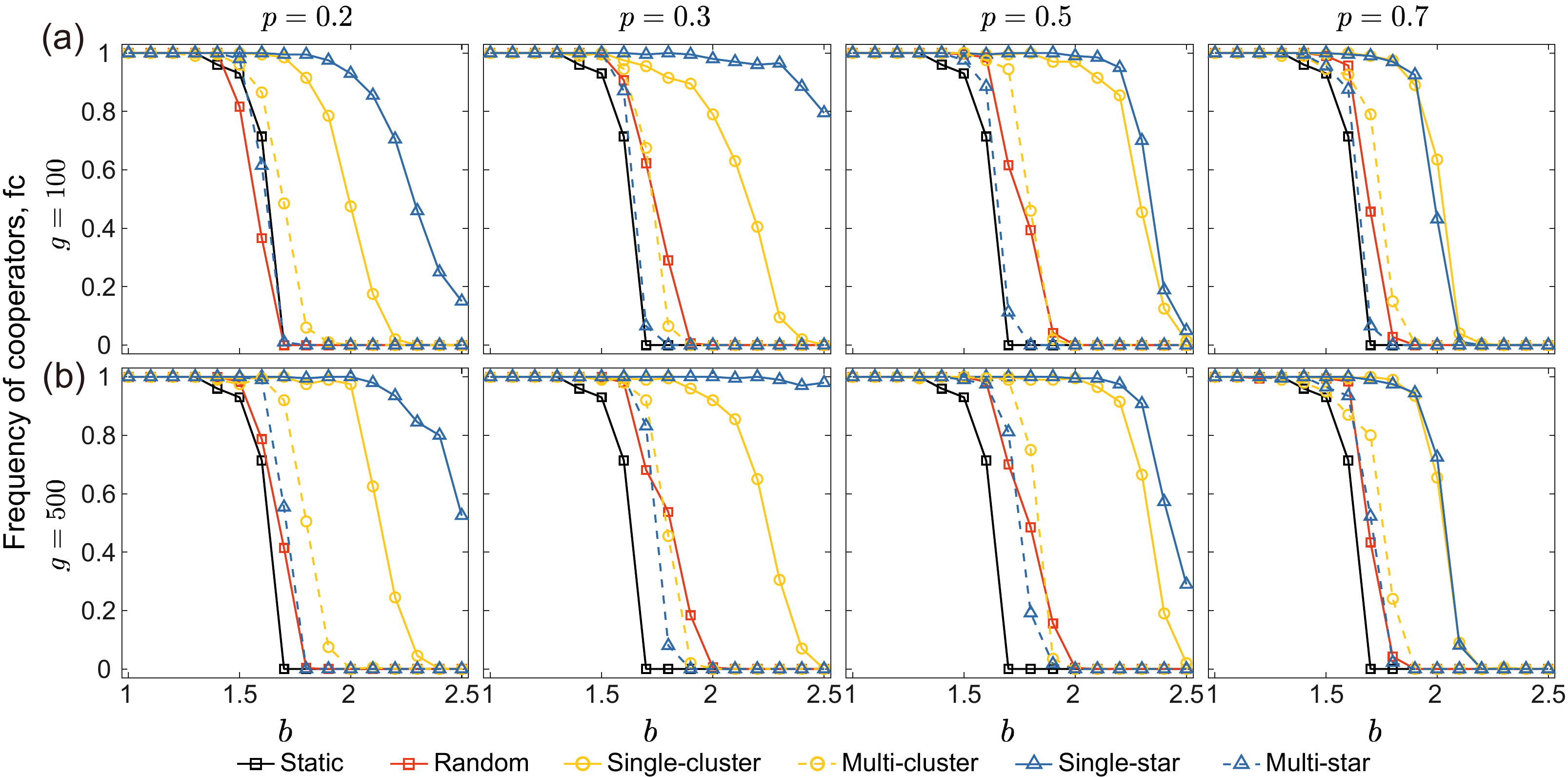}
                \caption{Structural temporal networks generally promote the evolution of cooperation. 
                We show the frequency of cooperation as a function of the temptation $b$ in the prisoner dilemma game
                on both static and structural temporal networks with different numbers of round ($g=100$ in ({a}), $500$ in ({b})) and active proportions $p=0.2, 0.3, 0.5, 0.7$ from left to right.
                We show that single-star temporal networks (blue triangle, solid line) have the highest frequency of cooperation, followed by the single-cluster temporal networks (yellow circle, solid line), while the multi-star temporal networks (blue triangle, dashed line) and multi-cluster temporal networks (yellow circle, dashed line) present only slight enhancement of fc compared to the random counterpart (red square, solid line). In contrast, the static network (black square, solid line) is the least effective structure in promoting cooperation.
                We calculate the equilibrium frequency of cooperation by averaging the results of $2\times 10^3$ rounds after a transient time of $5\times 10^4$ generations for $200$ independent realizations of the same network. 
                Parameter settings: network size $N=400$, average degree $k=8$, snapshot number $M=200$, selection intensity $s=2$.
                }\label{Fig2}
            \end{figure*}

            We first confirm that temporal networks can facilitate the emergence of cooperation compared to static networks (Fig.~\ref{Fig2}). The four temporal networks with constructed structures generally promote the emergence of cooperation more than random temporal networks. Interestingly, the single-star structure has the highest frequency of cooperators, which presents remarkable performance compared to all other structures in most cases. Specifically, this structure exhibits the strongest promotion of cooperation when the proportion of active edges ($p$) is set to 0.3, and almost full cooperation can be maintained even when facing a high temptation to defect ($b$). The single-cluster structure ranks second in promoting the emergence of cooperation. In contrast, multi-star and multi-cluster structures show almost no increase in the frequency of cooperators compared to random temporal networks. {Furthermore, we show that despite the single-star structure presenting a higher ability to facilitate cooperation compared to the single-cluster counterpart, the multi-cluster structure performs better than the multi-star structure in all cases}.
            \begin{figure*}[htbp]
                \centering
                \includegraphics[width=200pt]{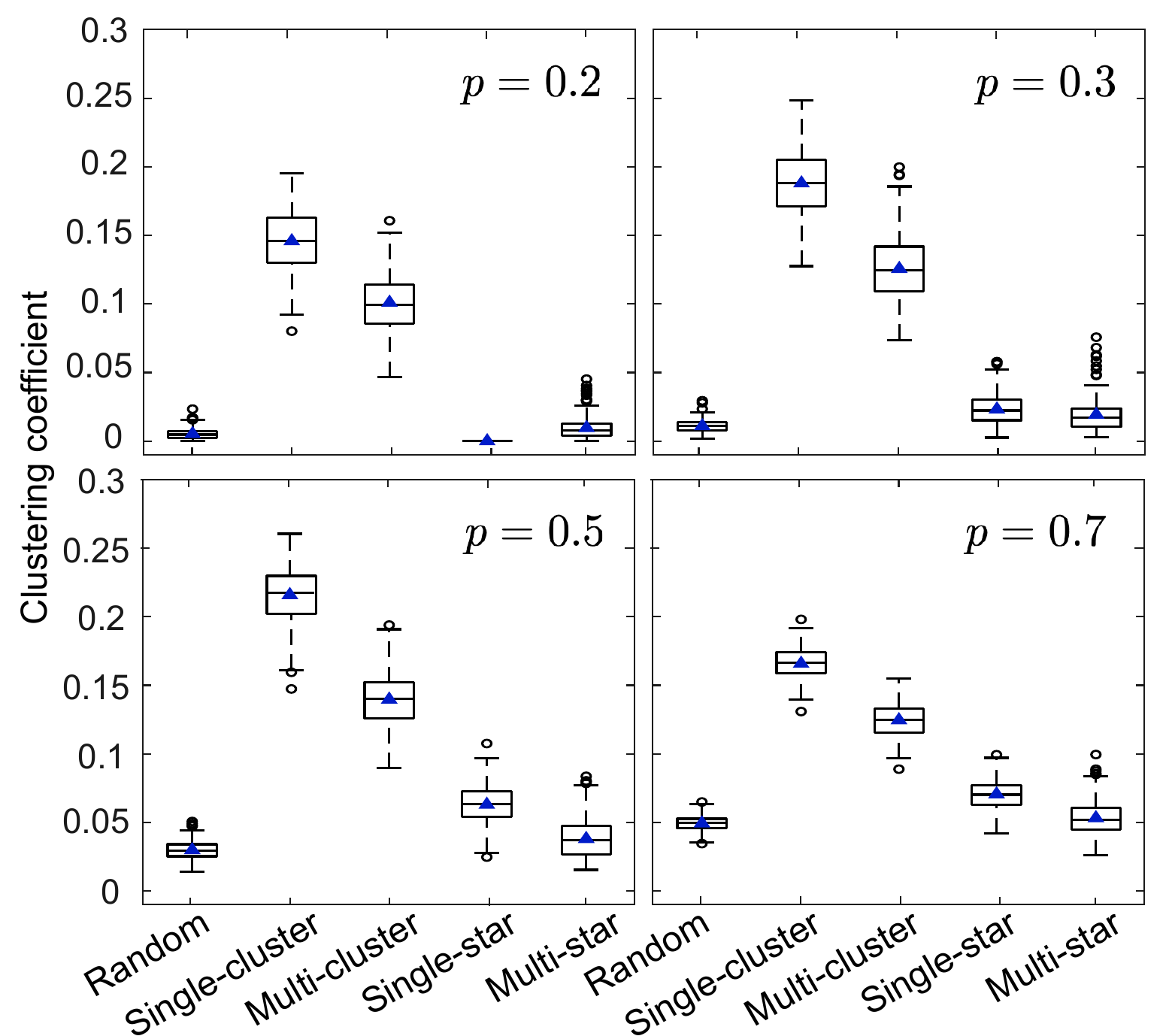}
                \caption{Clustering coefficients for different structural temporal networks. We show the values of the clustering coefficients of the temporal networks with a box plot, where the triangles denote the mean values of the clustering coefficients. We show that the single-cluster temporal networks have the highest clustering coefficient, followed by the multi-cluster temporal networks. However, the clustering coefficients of the single-star temporal networks and multi-star temporal networks are very close to those of random structure.}
                \label{Fig3}
            \end{figure*}

            Surprisingly, we find that the clustering coefficient does not effectively capture its ability to promote cooperation in temporal networks, despite its significance in static networks. Figure \ref{Fig3} shows that clustering coefficients in temporal networks have a limited correlation with cooperative levels. The single-star structure exhibits a higher frequency of cooperation than the single-cluster structure (Fig.~\ref{Fig2}) but with a lower clustering coefficient. In contrast, the multi-cluster structure has a larger clustering coefficient and promotes cooperation better than the multi-star structure. 

        \subsection{Mechanism of promoting cooperation for the single-star temporal network}
        \begin{figure*}
            \centering
            \includegraphics[width=\linewidth]{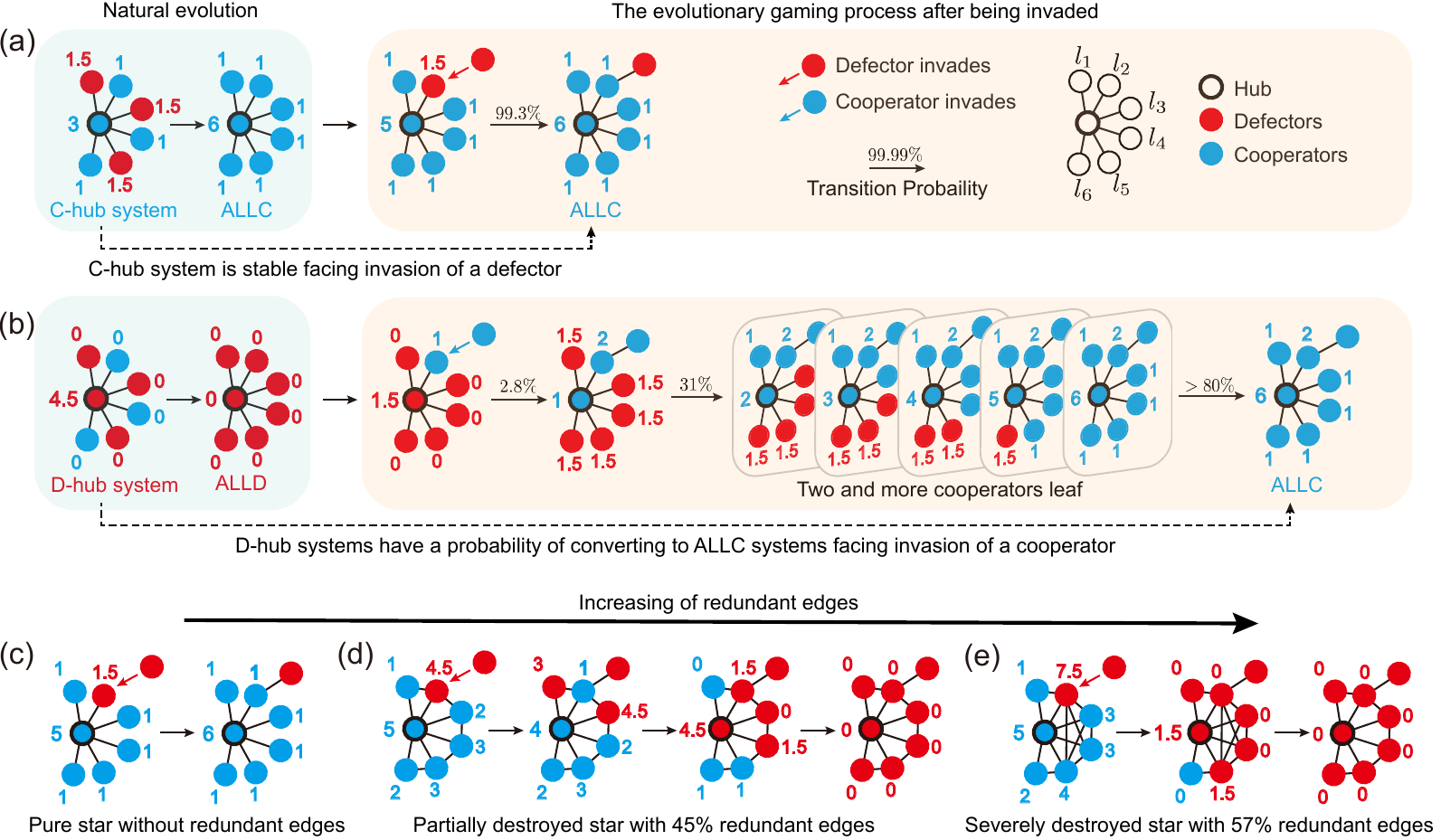}
            \caption{{Illustration on the mechanism of star structures to} facilitate cooperation. 
                We consider a system with a simple star structure, with each hub connected to six neighbours. 
                ({a}) When the hub is a cooperator, the system quickly switches to full cooperators and can easily resist invasion by defectors ($\pi>0.993$, see the detailed calculations of the transition probability in Supplementary Note 2 and Supplementary Fig.~S7).
                ({b}) When the hub is a defector, the system quickly switches full defectors, while it still has probability $\pi\approx 0.023$ of switching to full cooperators when facing the invasion of a cooperator.
                ({c}) We show that pure star structures with full cooperators can resist a defector's invasion.
                ({d}) When the star structure is partially destroyed, the resistance to the invasion of the defectors is reduced.
                ({e}) With the redundant edges increasing, the star structure is completely destroyed, and can not resist the invasion of the defectors.
                Here we choose the temptation $b = 1.5$ for all cases.
            }
            \label{FigS1}
        \end{figure*}
            To uncover the mechanism of temporal networks for promoting cooperation, we start by examining the single-star structure, which exhibits the most effective promotion of cooperation but has a low clustering coefficient. Given its abundance of star structures, we first investigate the propensity of stars to facilitate cooperation.
            Figure \ref{FigS1} shows the evolutionary dynamics of a simplified star population. For a hub with six neighbours, when half of the neighbours choose to cooperate and the other half defect, the strategy of the whole population will quickly shift to that of the hub. 
            After that, the defector invades the node $l_2$ of the population of full cooperators (ALLC, Fig.~\ref{FigS1}a), where the hub gains $P_{\mathrm{hub}}=5$ and the node $l_2$ gains $P_{l_2}=1.5$. According to the updating rule, the probability that the hub is invaded by a defector is
            \begin{equation}
                \begin{aligned}\label{eq1}
                    \pi_{\mathrm{hub}_{C\rightarrow D}}=\frac{1}{k_{\mathrm{hub}}}\cdot p_{\mathrm{hub}\rightarrow l_2}\approx0.00015,
                \end{aligned}
            \end{equation}
            where $p_{\mathrm{hub}\rightarrow l_2}$ denotes the probability that the hub node adopts its neighbour $l_2$'s strategy, and thus $\pi_{\mathrm{hub}_{C\rightarrow D}}$ is extremely low and the cluster of cooperators is stable to against the invasion.
            
            Similarly, for the population of full defectors (ALLD, Fig.~\ref{FigS1}b), the probability that the hub chooses to defect is
            \begin{equation}\label{eq2}
                \begin{aligned}
                    \pi_{\mathrm{hub}_{D\rightarrow C}}=\frac{1}{k_{\mathrm{hub}}}\cdot p_{\mathrm{hub}\rightarrow l_2}\approx0.045.
                \end{aligned}
            \end{equation}
            Comparing Equations \eqref{eq1} and \eqref{eq2}, it can be seen that $\pi_{\mathrm{hub}_{C\rightarrow D}}$ is much smaller than $\pi_{\mathrm{hub}_{D\rightarrow C}}$, thus the cluster of cooperators is more stable than that of defectors. After the defective cluster is invaded by cooperators, the probability of the cooperator driving one or more neighbours to cooperators while the hub can remain its strategy is
            \begin{equation}\label{eq3}
                \begin{aligned}
                    \pi_{\hspace{0.5mm}\mathrm{leaf}_{D\rightarrow C}}&=\left[\frac{1}{k_{\mathrm{hub}}}+\frac{k_{\mathrm{hub}}-1}{k_{\mathrm{hub}}}\cdot \left(1-p_{\mathrm{hub}\rightarrow l_1}\right)\right] \\&\cdot \left[1-\left(1-p_{l_1\rightarrow \mathrm{hub}}\right)^{k_{\mathrm{hub}}-1}\right]\approx 0.31.                    
                \end{aligned}
            \end{equation}
            Afterwards, this population will move towards ALLC with a high probability (large{r} than $0.80$, see Supplementary Note 2 for details, Supplementary Fig.~S7).

            In fact, Equations \eqref{eq1}, \eqref{eq2} and \eqref{eq3} are only one path for the ALLD population to be restored to the ALLC population after being invaded, and we can get the actual probability of 0.23 by Monte Carlo simulations (see Supplementary Note 2).
            In summary, the hub of the population with full cooperators is very hard to be invaded due to the very high payoffs obtained from its cooperative neighbours, and the population of full cooperators can easily invade most of the defective individuals, thus increasing the number of cooperators in the whole network. 
            From this perspective, a network without any triangle but with numerous star hubs interconnected is more conducive to the formation of clusters of cooperators. 

            We further find that as the proportion of edges selected from the underlying static network increases, the pure star structure undergoes a gradual disintegration. Figure \ref{FigS1}c-e illustrates the progressive deterioration of the star structure in panels a, b, and c. Alongside the gradual destruction of the star structure, defectors gain higher payoff on invasions, while the resistance of cooperative clusters diminishes gradually.
            This explanation can be verified in Fig.~\ref{Fig3} when $p=0.3$, each snapshot of the single-star structure activates all points, and the star structure has advantages compared to others, so it promotes cooperation best. As $p$ increases ($p>0.5$), the single-star structure gradually approximates a static scale-free network. The highly stable hubs with high-payoffs within the star, previously mutually reinforcing, transition into a stable cluster of cooperators constructed by hubs. Consequently, the effect of single-star structure on cooperation decreases rapidly.

            Our previous analyses are primarily based on a single snapshot. In a temporal network with a sequence of snapshots, the advantage of the star structure becomes more pronounced. As the game progresses, the number of cooperating nodes in the star-rich snapshot increases. When the system switches to a new snapshot, networks with more cooperative nodes have a greater chance of becoming the central nodes of stars in the new snapshot. This process reduces the difficulty for cooperators in static networks to invade the ALLD-star, as the structure of ALLD-star is naturally disrupted after the temporal switch. Therefore, compared to situations without snapshot switching, more ALLC-star structures will emerge. Over the course of the evolutionary process, the temporal network is more likely to reach a state of full cooperation than the corresponding static network.
        \subsection{Quantifying the effectiveness of the characteristics of temporal networks in fostering cooperation}
            The edges of the network define the interaction between individuals {and serve as pathways for strategy transmission}, having an important role in the emergence of collective cooperation. Inspired by the superiority of single-star structure in promoting cooperation, which has a large number of edges between large-degree nodes and that between large-degree and small-degree nodes, we suspect that edges between small-degree nodes are not conducive to cooperation. Therefore, we next propose a metric based on this intuition to evaluate the ability of the dynamic network structure to facilitate cooperation.

            \begin{figure*}
                \centering
                \includegraphics[width=\linewidth]{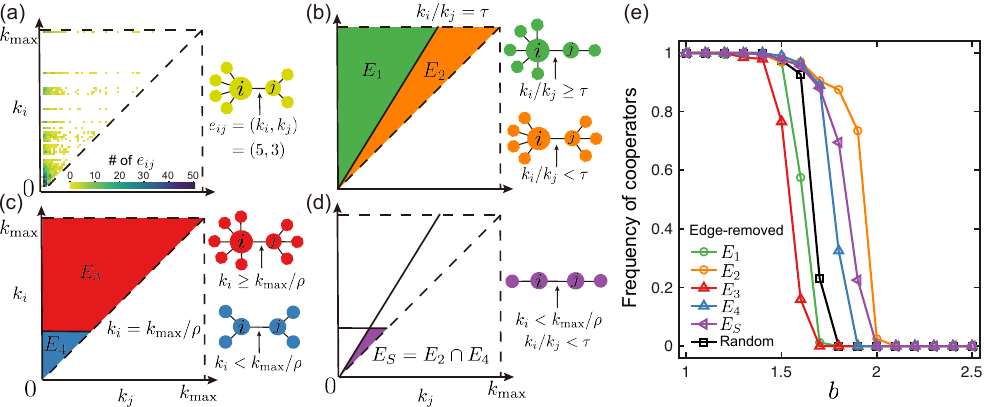}
                \caption{The classification of the capacity of different edge types on promoting cooperation.
                ({a}) We show the frequency of edge pairs with degree $k_i$ and $k_j$ in a heatmap. Given that this is an undirected graph, the edges we count are considered to satisfy $k_i\ge k_j$, and the colours from light yellow to dark blue indicate the frequency of edges.
                (b) We categorize the types of edges according to the size of $k_i/k_j$. The region $E_1$ (green) indicates edges with significant degree disparities at their ends, while $E_2$ (orange) encompasses edges with minor degree differences.
                (c) We further categorize the types of edges according to the size of $k_i$, where $E_3$ (red) signifies the set of edges with larger end degrees, and $E_4$ (blue) includes edges with smaller degrees.
                (d) By combining the way of classifying edges in ({b}) and ({c}), we select the intersection of $E_2$ and $E_4$---the set $E_S$ (purple)---as our final criterion.
                (e) We run the evolutionary game on the networks with the same number of removed different types of edges compared to the network with randomly removed edges. 
                Here we choose $\tau=1.3$ and $\rho=5$ to distinguish the effect of removing $E_S$ from other cases. The robustness of our classification methods in Supplementary Fig.~S8. Other parameters are the same as those in Fig.~\ref{Fig2}.
                }\label{Fig5}
            \end{figure*}

            We count the number of different types of edges of the underlying static network in Fig.~\ref{Fig5}a. The edge between node $i$ and $j$ is categorized by the relative size $k_i/k_j$ (Fig.~\ref{Fig5}b, here $k_i$ denotes the degree of node $i$) and absolute size $k_i$ (Fig.~\ref{Fig5}c, assuming $k_i>k_j$). In this way, the former allows us to observe the impact of edges between large-degree to small-degree nodes, while the latter examines the effects of edges between large-degree nodes. The outcome of the category can be expressed by 
            $$\begin{aligned}
                E_1&=\left\{e_{ij}: v_i, v_j\in V, e_{ij}\in E, \frac{{k}_{i}}{{k}_{j}} \ge \tau\right\}\\
                E_2&=\left\{e_{ij}: v_i, v_j\in V, e_{ij}\in E, \frac{{k}_{i}}{{k}_{j}} < \tau\right\}\\
                E_3&=\left\{e_{ij}: v_i, v_j\in V, e_{ij}\in E, {k}_{i} \ge \frac{k_{\max}}{\rho}\right\}\\ 
                E_4&=\left\{e_{ij}: v_i, v_j\in V, e_{ij}\in E, {k}_{i} < \frac{k_{\max}}{\rho}\right\},
            \end{aligned}$$
            where $V$ is the set of nodes of the network, $E$ is the set of edges, $\tau$ controls the ratio between $k_i$ and $k_j$ and $\rho$ controls the magnitude of $k_i$ relative to $k_{\max}$. 
            In order to confirm the impact of different edges on cooperation, we remove various edge sets for the underlying static network and perform evolutionary games to determine the frequency of cooperation, and compare the results with those of a network that is randomly removed by the same number of edges.

            \begin{figure*}
                \centering
                \includegraphics[width=\linewidth]{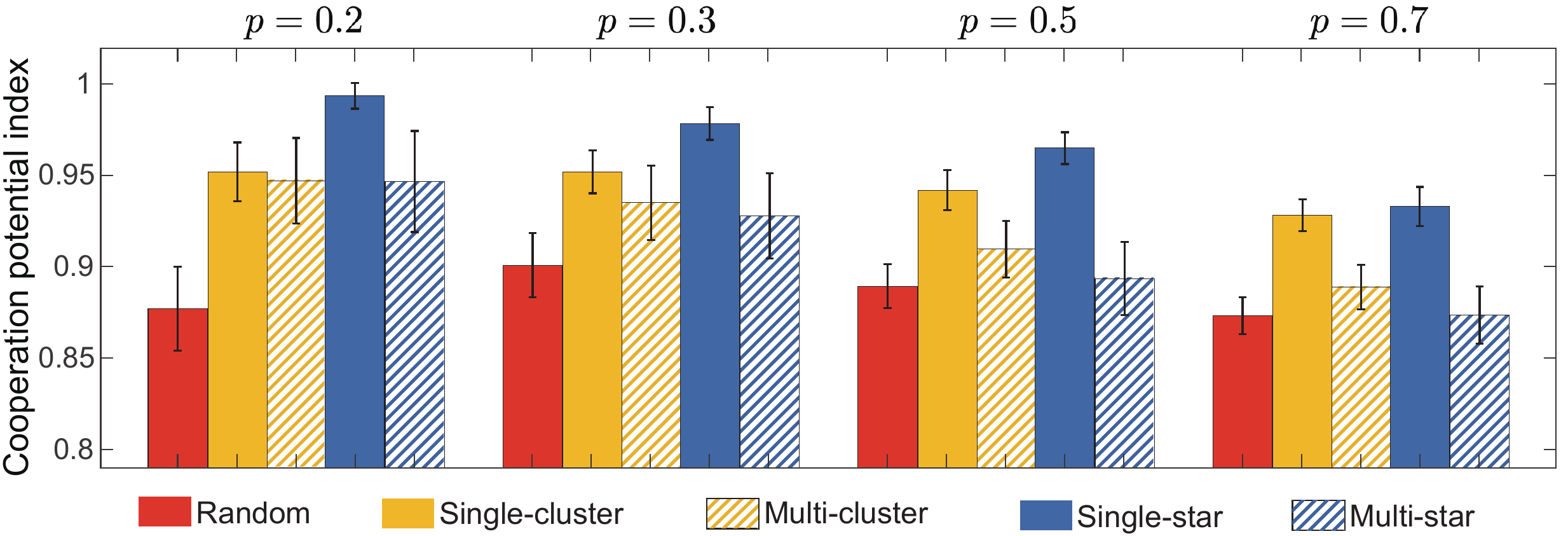}
                \caption{Evaluating the efficiency of networks in promoting cooperation through the CPI. 
                For temporal networks with different edge selection proportions $p$, we show the CPIs counted from 200 snapshots and plot the error bar for each type of network at $p=0.2,0.3,0.5$ and $0.7$. 
                Our results reveal that the single-star structure (solid blue) exhibits the highest CPI, followed by the single-cluster temporal network (solid yellow). The multi-cluster (striped yellow) and multi-star temporal network (striped blue) display comparable CPIs, with the former marginally outperforming the latter. In contrast, the random structure is characterized by the lowest CPI, suggesting its relatively inferior effectiveness in promoting cooperation.
                Generally, the relative magnitude of CPI values is consistent with the differences in the frequency of cooperation shown in Fig.~\ref{Fig2}.
                }\label{Fig7}
            \end{figure*}

            We show the ability of the network to promote cooperation is enhanced, after removing the edges in the $E_2$ or $E_4$ regions in Fig.~\ref{Fig5}e. This implies that both regions are not conducive to cooperation. Therefore, we take their intersection $E_S$ (Fig.~\ref{Fig5}d) as the set of key edges that inhibit cooperation, which is defined by  
            $$E_S=E_3\cap E_4.$$
            All the edges $e_{ij}^S$ in the $E_S$ contain small $k_i$ and small $k_i/k_j$, which have a substantial inhibitory effect on cooperation (Fig.~\ref{Fig5}e). In particular, the network with the set of $E_S$ edges removed presents an edge distribution that closely mirrors that of a single-star structure in terms of the type of edges. Therefore, we utilize the proportion of edges within the $E_S$-region to construct the Cooperation Potential Index (CPI) by
            $$\mathrm{CPI} = 1 - \frac{|E_S|}{|E|},$$
            which assesses the network structure's capacity in promoting cooperation (Supplementary Fig.~S9). We demonstrate the CPI values of different network structures (Fig.~\ref{Fig7}), representing the relationship between the magnitude of their ability to facilitate cooperation. 
            Additionally, as the edge selection proportion increases, it is noteworthy that the cooperative abilities of all networks experience a certain degree of decline, in line with the findings presented in Fig.~\ref{Fig2}. Thus, within the current framework, the CPI can be used as an effective tool to measure the cooperation potential of structural temporal networks. The CPI here is robust to changes in the heterogeneity and average degree of the scale-free networks, and for more results, see Supplementary Figs.~S10-S13.

    \section{DISCUSSION}
        We have explored the effect of different characteristics of temporal networks on collective cooperation, and here we uncover that they can generally promote the emergence of cooperation, due to the advantages presented by the network structure (or the number of specific edges) rather than high clustering coefficients previously discovered based on static networks. Additionally, by classifying edges on the network, we propose a simple metric, CPI, which qualitatively quantifies the ability of different network structures on facilitating collective cooperation. 
        
        A natural application based on our findings includes exploring and designing the structure of higher-order graphs. In reality, multiple nodes often interact simultaneously, represented by hypergraphs or simplicial complexes \cite{majhi2022dynamics,yang2024generalized}. Interactions between groups in higher-order networks are more complex than in paired interactions \cite{gokhale2010evolutionary, wang2024networked}. The previous study shows temporal hypernetworks may promote collective cooperation \cite{230601300}, our findings can couple hypernetworks with network structures to explore the effect of hyperedge structure on collective cooperation, and further design optimal temporal hypernetworks that promote collective cooperation, all of which will be the meaningful focus of future research.

        In the realm of nature, the learning capabilities of individuals are not uniform, which is reflected in research as each individual has distinct opportunities for strategy updates. Our present investigation focuses on that individuals have equal chances for strategy updates. Nevertheless, recent research suggests that the heterogeneous updating opportunities for individual strategies can significantly impact the effectiveness of different network structures in promoting cooperation \cite{meng2024dynamics}. Consequently, in scenarios where individuals exhibit different update speeds, or more broadly, when they have varying information \cite{wang2023imitation} as well as different behavioural traits \cite{zhou2024beyond} during strategy updates, the impact of the temporal network structure on cooperation becomes a central area of research. Specifically, identifying a more unified structural framework could provide valuable insights regarding the design of the optimal structure of temporal networks.
        
        Our framework provides an effective way to optimize the structure of temporal networks to improve collective cooperation. For example, we can construct temporal networks with a single-star structure by altering the timing of individuals' interactions, which can be applied in engineering systems to enhance autonomous collaboration among agents. The understanding and application of the dynamics of temporal networks thus not only provide valuable insights into the behaviour of multi-agent systems but also offer practical strategies for designing interventions that promote autonomous collaboration. Our findings can be utilized to achieve more efficient swarm intelligence, harmonious social interaction for collective benefits and even sustainable economic growth.
        
    \section{Appendix}
        \subsection{Modelling process of different structural temporal networks}

            The construction process of the temporal network is briefly described herein. Detailed pseudocode is available in the Supplementary Information.
            
            \subsubsection{Random temporal network:}
                Randomly select $p$-proportional edges on the underlying static network $G(V, E)$.
            \subsubsection{Single-cluster temporal network:}
                \begin{enumerate}
                    \item At each time step $t$, a node $i$ is randomly selected on the underlying static network $G(V, E)$. 
                    \item Initiating from node $i$, a Breadth-First Search (BFS) algorithm is executed. Sequentially, every edge encountered in the BFS traversal is added into the temporal network $G_S(V_S, E_S)$. This process continues until $|E_S|=p*|E|$.
                    \item If $|E_S|<p*|E|$, repeat steps 1-3.
                \end{enumerate}
            \subsubsection{Multi-cluster temporal network:}
                \begin{enumerate}
                    \item At each time step $t$, a node $i$ is randomly selected on the underlying static network $G(V, E)$.
                    \item For each neighbour $j$ of node $i$, add the edge $e_{ij}$ to the current temporal network $G_S(V_S, E_S)$, until $|E_S|=p*|E|$.
                    \item For each pair of neighbours $j$ and $k$ of node $i$, add the edge $e_{ij}$ to $G_S$ (given $e_{ij}\in G$), until $|E_S|=p*|E|$.
                    \item If $|E_S|<p*|E|$, repeat steps 1-4.
                \end{enumerate}
            \subsubsection{Single-star temporal network:}
                \begin{enumerate}
                    \item At each time step $t$, randomly select one node $i$ to be the centre node. Mark node $i$ as active.
                    \item Initiating from node $i$, a Breadth-First spanning tree (BFSTree) algorithm is executed. Sequentially, every edge encountered in the BFSTree traversal is added into the temporal network $G_S(V_S, E_S)$. This process continues until $|E_S|=p*|E|$.
                    \item If $|E_S|<p*|E|$, repeat steps 1-3.
                \end{enumerate}
            \subsubsection{Multi-star temporal network:}
                \begin{enumerate}
                    \item At each time step $t$, a node $i$ is randomly selected on the underlying static network $G(V, E)$.
                    \item For each neighbour $j$ of node $i$, add the edge $e_{ij}$ to the current temporal network $G_S(V_S,E_S)$, until $|E_S|=p*|E|$.
                    \item If $|E_S|<p*|E|$, repeat steps 1-3.
                \end{enumerate}

\bibliography{../reference/bib1}
\end{document}